\documentstyle[aps,preprint]{revtex}

\begin{document}

\draft

\title
{\bf Self-Consistent Theory of Metal-Insulator Transitions in
Disordered Systems}
\author{E.Z.Kuchinskii,\ V.V.Sadovskii,\ V.G.Suvorov,\ M.A.Erkabaev}
\address
{Institute for Electrophysics, Russian Academy of Sciences, \\
Ural Branch,\ Ekaterinburg 620219,\ Russia\\
E-mail: sadovski@ief.e-burg.su}
\maketitle

\vskip 1.5cm
\begin{center}
\sl{Submitted to JETP, November 1994}
\end{center}

\begin{abstract}

Self-consistent theory of electron localization in disordered systems is
generalized for the case of interacting electrons. We propose and
critically compare a number of possible self-consistency schemes which take
into account the lowest perturbation theory contributions over the
interaction. Depending on self-consistency scheme we can obtain either the
continuous metal-insulator transition or that with the minimal metallic
conductivity. Within the continuous transition approach we calculate the
frequency dependence of generalized diffusion coefficient both for metallic
and insulating phases. We also consider interaction renormalization of the
single-electron density of states which demonstrates the growth of the
effective pseudogap as system goes from metal to insulator.

\end{abstract}

\newpage
\narrowtext

\section{Introduction}

Theoretical description of disorder-induced metal-insulator transitions
constitutes one of the major problems of condensed matter theory [1,2].
The main difficulties here are connected with consistent account of
interelectron interactions, the major importance of which was already
demonstrated even in case of weakly disordered metals [3]. In recent years
this problem was actively studied within the renormalization group approach
[1,2,4], generalizing the usual scaling theory of localization [1].
Though very successful in many respects this approach was not able to produce
a complete solution of the problem and in particular it almost failed in its
ability to predict the concrete and experimentally verifiable dependences
of physical properties on parameters of the system which control the
transition. Renormalization group (scaling) approach is well fitted for the
analysis of the region very close to transition itself [2], while
experimentally
we are usually dealing with variation of physical properties in some wide
region of parameters controlling the transition. In particular, up to now
there are no papers treating the insulating phase within renormalization group
approach accounting for interactions.

In the theory of metal-insulator transitions neglecting the interactions along
with scaling approach [1], the so called self-consistent theory of localization
is widely used [5,6,7] and provides a rather successful interpolation scheme,
which allows calculations of the main physical properties of a system for the
wide region of parameters from weakly disordered metal to the Anderson
insulator. This theory also reproduces all the major results of the scaling
approach [5,6]. While the rigorous diagrammatic procedure leading to the
self-consistent scheme is still unknown, this approach leads to results which
are in quantitative accordance with exact numerical data modelling the
Anderson transition [8,9]. First attempts to account for the interaction
effects within the self-consistent theory of localization were undertaken in
Refs.[10,11] (Cf. also Ref [6]).

This paper attempts to formulate some version of self-consistent theory of
disorder-induced metal-insulator transition accounting for the first-order
perturbation theory corrections over interelectron interactions. The
difference with the previous attempts [10,11] is that we shall try to take
into account the interaction effects upon the generalized diffusion
coefficient, which represents the main physical characteristic to be
determined self-consistently. Unfortunately, we can propose a number of
alternative formulations of self-consistency which lead to different
predictions concerning the physical properties. We can only choose between
these different schemes only using some guidance from the experimental
behavior during metal-insulator transition. In particular, depending upon the
self-consistency scheme we can get either the continuous metal-insulator
transition or that with the so called minimal metallic conductivity.
Modern experiments lead to the picture of continuous transition [1,2]. In the
framework of such a scheme we shall calculate the frequency dependence of
generalized diffusion coefficient as well as renormalization of single-electron
(tunneling) density of states due to interactions for the wide region of
parameters varying during the metal-insulator transition.

\section{Interaction corrections to conductivity close to the
Anderson transition.}

Let us introduce two-particle Green's function for electrons in the random
field of impurities [5]:

\begin{equation}
\Phi ^{RA}_{{\bf p p'}}({\bf q}\omega\varepsilon)=<G^{R}({\bf p_{+}p'_{+}}
\varepsilon+\omega)G^{A}({\bf p_{-} p'_{-}}\varepsilon)>
\end{equation}
where ${\bf p_{\pm}}={\bf p} \pm \frac{ {\bf q} }{2}$,\ and  $<\ldots >$
- denotes the averaging over the random potential.

Let us introduce the appropriate full vertex-part
$\Gamma^{RA}_{{\bf pp'}}({\bf q}\omega)$
and also the "triangular" vertex $\gamma^{RA}({\bf pq}\omega)$ (Cf. Fig.1).

Within self-consistent theory of localization the full vertex-part and
triangular vertex are given by [6]:

\begin{equation}
\Gamma^{RA}_{\bf pp'}({\bf q}\omega)=\frac{2\gamma\rho U^{2}}{-i\omega
+D(\omega)q^{2}}
\label{G}
\end{equation}

\begin{equation}
\gamma^{RA}({\bf pq}\omega)=\frac{2\gamma}{-i\omega +D(\omega)q^{2}}
\label{g}
\end{equation}
where $\gamma=\frac{1}{2\tau}=\pi\rho U^2 N_{0}(0)$ - is the usual Born
decay rate, $\rho$ - impurity concentration, $U$ - impurity potential,
$N_{0}(0)$ - density of states at the Fermi level for noninteracting electrons,
$D(\omega)$ - the generalized diffusion coefficient.

Expressions for the vertex parts actually coincide with those obtained in the
"ladder" approximation with the usual Drude-like diffusion coefficient $D_{0}$
replaced by frequency-dependent generalized diffusion coefficient $D(\omega)$,
defined from the solution of the following self-consistent equation [5,7]:

\begin{equation}
\frac{D_{0}}{D(\omega)}=1+ \frac{1}{ \pi N_{0}(0) } \sum_{|{\bf q}|<k_{0}}
\frac{1}{-i\omega + D(\omega)q^2}
\label{ES}
\end{equation}
where $k_{0}=min\{l^{-1},k_{F}\}$ - is integration cutoff momentum, $l$ -
is the mean-free path, $k_{F}$ - is Fermi momentum.

In three-dimensional system the generalized diffusion coefficient takes the
following form:

\begin{equation}
\frac{D_{s}(\omega)}{D_{0}}= \left\{\begin{array}{lll}
\frac{3\pi\lambda}{2}{\left(\frac{\omega_{c}}{E_{F}}\right)}^{\frac{1}{3}}
=\alpha &
\quad \omega \ll \omega_{c}, \quad \alpha >0  & \quad \mbox{Metal}   \\
\frac{3\pi\lambda}{2}{\left(\frac{-i\omega}{E_{F}}\right)}^{\frac{1}{3}} &
\quad \omega \gg \omega_{c} & \quad \mbox{Metal and Insulator} \\
\frac{3\pi\lambda}{2}{\left(\frac{\omega_{c}}{E_{F}}\right)}^{-\frac{2}{3}}
\left(\frac{-i\omega}{E_{F}}\right) &
\quad \omega \ll \omega_{c}, \quad \alpha <0 & \quad \mbox{Insulator}
\end{array}\right.
\label{Ds}
\end{equation}
where $\lambda=\frac{\gamma}{\pi E_{F}}$ - is dimensionless disorder parameter,
$E_{F}$ - Fermi energy, $\omega_{c}=\left( {\frac{2|\alpha|}{3\pi\lambda}}
\right)^{3}E_{F}$ - is the characteristic frequency,
$\alpha=1-3\lambda x_{0}$ - is the parameter which controls metal-insulator
transition, $x_{0}=\frac{k_{0}}{k_{F}}$ - the dimensionless cutoff.

Neglecting localization contributions first-order corrections to conductivity
due to Coulomb interaction are determined by diagrams shown in Fig.2 [12].
It was shown in Ref.[12] that the total contribution of diagrams (a), (b) and
(c) is actually zero and conductivity correction reduces to that determined by
diagrams (d) and (e). Here we neglect also the so called Hartree corrections to
conductivity [3,12], which is valid in the limit of $2k_{F}/\kappa \gg 1$,
where $\kappa$ - is the inverse screening length. This inequality, strictly
speaking, is valid for systems with low electronic density, which are most
interesting for experimental studies of disorder induced metal-insulator
transitions. Also, if we remember the known results on the divergence of
screening length at the metal-insulator transition, we can guess
that this approximation becomes better as we approach the transition.
The point-like interaction model used below has to be understood in this sense.

Using the explicit form of impurity vertexes (\ref{G}) and (\ref{g}),
we obtain the following correction to conductivity due to interactions
(Cf. [12]):

\begin{equation}
\delta \sigma (\omega)=\frac{32i}{\pi d}e^{2}N_{0}(0)D_{0}^{2}
\int\limits_{\omega}^{\infty}\frac{d\Omega}{2\pi}\int \frac{d^{d}q}{(2\pi)^{d}}
\frac{V(q\Omega)q^{2}}{(-i(\Omega +\omega)+D(\Omega +\omega)q^{2})
(-i\Omega+D(\Omega)q^{2})^{2}}
\label{dG}
\end{equation}
HereØ $e$ - is the electronic charge, $d$ - is spatial dimensionality,
$V(q\Omega)$ - interaction. For simplicity in the following we shall mainly
consider the point-like interaction $V(q\Omega)=V_{0}$.
In case of dynamically screened Coulomb interaction at small $q$ and $\Omega$
we have [3]:
\begin{equation}
V(q\Omega)=\frac{1}{2N_{0}(0)}\frac{-i\Omega +D(\Omega)q^{2}}{D(\Omega)q^{2}}
\label{V}
\end{equation}

Then for the first-order correction to diffusion from the point-like
interactions we obtain:
\begin{equation}
\frac{\delta D_{c}(\omega)}{D_{0}}=\frac{\delta \sigma(\omega)}
{2e^{2}N_{0}(0)D_{0}}=
\label{dD1}
\end{equation}
$$=\frac{8i}{\pi d}\mu D_{0}\frac{1}{\pi N_{0}(0)}
\int\limits_{\omega}^{\infty}d\Omega\int \frac{d^{d}q}{(2\pi)^{d}}
\frac{q^{2}}{(-i(\Omega +\omega)+D(\Omega +\omega)q^{2})(-i\Omega+
D(\Omega)q^{2})^{2}}$$
where $\mu=V_{0}N_{0}(0)$.

Using in (\ref{dD1}) the form of diffusion coefficient obtained in
self-consistent theory of localization (\ref{Ds}) in three-dimensional case
we obtain:

\begin{equation}
\frac{\delta D_{cs}(\omega)}{D_{0}}= -\mu\frac{2}{3\pi\lambda}\left\{
\begin{array}{lll}
\frac{3\sqrt{3}+2\sqrt{2}}{2\pi}{\left(\frac{E_{F}}
{\omega_{c}}\right)}^{\frac{1}{3}} &
\quad \omega \ll \omega_{c}, \quad \alpha >0  & \quad \mbox{Metal}   \\
\frac{3}{\pi}{\left(\frac{E_{F}}{-i\omega}\right)}^{\frac{1}{3}} &
\quad \omega \gg \omega_{c} & \quad \mbox{Metal and Insulator} \\
\frac{1}{\pi}{\left(\frac{\omega_{c}}{E_{F}}\right)}^{\frac{5}{3}}
{\left(\frac{E_{F}}{-i\omega}\right)}^{2} &
\quad \omega \ll \omega_{c}, \quad \alpha <0 & \quad \mbox{Insulator}
\end{array}\right.
\label{dD}
\end{equation}
It is easy to see that the correction to diffusion coefficient (conductivity)
diverges as system approaches the Anderson transition from metallic phase
($\omega_{c}\rightarrow 0$), while in insulating phase divergence appears as
$\omega\rightarrow 0$.

If instead of the point-like interaction we take dynamically screened Coulomb
interaction (\ref{V}), correction to diffusion coefficient takes the form:

\begin{equation}
\frac{\delta D_{c}(\omega)}{D_{0}}=
i\frac{4\lambda}{\pi}\frac{D_{0}^{2}}{k_{F}}
\int\limits_{\omega}^{\infty}d\Omega\int\limits_{0}^{k_{0}} dq
\frac{q^{d-1}}{(-i(\Omega +\omega)+D(\Omega +\omega)q^{2})(-i\Omega+
D(\Omega)q^{2})D(\Omega)}
\label{dD2}
\end{equation}
Using here the form of diffusion coefficient from self-consistent theory
localization (\ref{Ds}), we get the correction like (\ref{dD}),
but with $\mu=\frac{1}{3}$. That is, the use of dynamically screened Coulomb
interaction instead of point-like leads just to a replacement of $\mu$
by a constant of order of unity and in the following we always assume
the point-like interaction.

Thus we can see that the correction to diffusion coefficient close to the
Anderson transition becomes much larger than diffusion coefficient of the
self-consistent theory itself. This obviously leads to the necessity to
formulate some kind of new self-consistency scheme taking interelectron
interaction into account from the very beginning.

\section{Self-consistency schemes.}

Let us start with the usual self-consistent theory of localization in the
absence of Coulomb interaction [5---7]. In case of weak disorder conductivity
is determined by Drude diffusion coefficient $D_{0}$. Summation of
"maximally crossed" diagrams leads to the following localization
correction to diffusion coefficient [5]:

\begin{equation}
\frac{\delta D(\omega)}{D_{0}}=-\frac{1}{ \pi N_{0}(0) }
\sum_{|{\bf q}|<k_{0}} \frac{1}{-i\omega + D_{0}q^2}
\label{dDLN}
\end{equation}
Now we can introduce the so called relaxation kernel $M(\omega)$,
which is connected with the generalized diffusion coefficient by the following
relation [5]:

\begin{equation}
M(\omega)=i\frac{2E_{F}}{dm}\frac{1}{D(\omega)}
\label{M}
\end{equation}
In particular, the Drude relaxation kernel is given by
$M_{0}=i\frac{2E_{F}}{dm}\frac{1}{D_{0}}=2i\gamma$.

Correction to relaxation kernel can be expressed via the correction to
diffusion coefficient as:

\begin{equation}
\delta M(\omega)=-i\frac{2E_{F}}{dm}\frac{\delta D(\omega)}{D(\omega)^{2}}=
-\frac{M(\omega)}{D(\omega)}\delta D(\omega)
\label{dM}
\end{equation}
Consider the usual Drude metal as the zeroth approximation:

\begin{equation}
\delta M(\omega)=-\frac{M_{0}}{D_{0}}\delta D(\omega)
\label{dM0}
\end{equation}
Replacing Drude diffusion coefficient $D_{0}$ in the diffusion pole of Eq.
(\ref{dDLN}) by the generalized one $D(\omega)$, and using this relation in
Eq.(\ref{dM0}), we obtain the main equation of self-consistent theory of
localization:

\begin{equation}
M(\omega)=M_{0}+\delta M(\omega) =M_{0}\left\{1+ \frac{1}{ \pi N_{0}(0) }
\sum_{|{\bf q}|<k_{0}} \frac{1}{-i\omega + D(\omega)q^2}\right\}
\label{Ms}
\end{equation}
which coincides with Eq.(\ref{Ds}) if we take into account that
$\frac{M(\omega)}{M_{0}}=\frac{D_{0}}{D(\omega)}$.

The first order corrections due to Coulomb interaction are determined by
diagrams (d) and (e) shown in Fig.2. Unfortunately we cannot propose the
uniquely defined self-consistency scheme which take into account corrections
due to interelectron interactions. We can only formulate several alternative
schemes and try to choose between them using some additional qualitative
guesses.

{\bf Scheme I.}\ \ Let us first use the approach similar to the usual
self-consistent theory of localization. As a zeroth approximation we take
Drude metal and consider localization and Coulomb corrections on equal footing
and introducing the self-consistency condition in terms of relaxation kernel.

In this case the relaxation kernel takes the following form:

\begin{equation}
M(\omega)=M_{0}+\delta M(\omega)
\label{Md}
\end{equation}
where $\delta M(\omega)=\delta M_{l}(\omega)+\delta M_{c}(\omega)=
-\frac{M_{0}}{D_{0}}(\delta D_{l}(\omega)+\delta D_{c}(\omega))$. Here the
localization correction to diffusion coefficient $D_{l}(\omega)$ is defined by
Eq.(\ref{dDLN}), while the Coulomb correction $D_{c}(\omega)$ is given by
Eq.(\ref{dD1}). Self-consistency procedure is reduced to the replacement of
$D_{0}$ by the generalized diffusion coefficient in all diffusion denominators.
As a result we obtain the following integral equation for the generalized
diffusion equation:

\begin{equation}
\frac{D_{0}}{D(\omega)}=  1+ \frac{1}{ \pi N_{0}(0) }\int \frac{d^{d}q}
{(2\pi)^{d}} \frac{1}{-i\omega + D(\omega)q^2}-
\label{I}
\end{equation}
$$-\frac{8i}{\pi d}\mu D_{0}\frac{1}{\pi N_{0}(0)}
\int\limits_{\omega}^{\infty}d\Omega\int \frac{d^{d}q}{(2\pi)^{d}}
\frac{q^{2}}{(-i(\Omega +\omega)+D(\Omega +\omega)q^{2})(-i\Omega+
D(\Omega)q^{2})^{2}}$$

{\bf Scheme II.}\ \ As a zeroth approximation we can now take the "dirty" metal
described by the usual self-consistent theory of localization and consider
first-order Coulomb corrections to this state. Self-consistency again is
realized through corrections to relaxation kernel.

Usual self-consistent theory is based upon Eq.(\ref{Ms}). Now we have to add
into the right-hand side of this equation the Coulomb correction to relaxation
kernel $\delta M_{c}(\omega)$. This correction has the following form:

\begin{equation}
\delta M_{c}(\omega)=-\frac{M_{s}(\omega)}{D_{s}(\omega)}\delta D_{c}(\omega)
\label{dMs}
\end{equation}
where $M_{s}(\omega)$ and $D_{s}(\omega)$ - are the relaxation kernel and
diffusion coefficient of the usual self-consistent theory of localization
 (\ref{Ds}).

In this case, the equation determining for diffusion coefficient reduces to:

\begin{equation}
\frac{D_{0}}{D(\omega)}=  1+ \frac{1}{ \pi N_{0}(0) }\int \frac{d^{d}q}
{(2\pi)^{d}} \frac{1}{-i\omega + D(\omega)q^2}-
\label{II}
\end{equation}
$$-\left(\frac{D_{0}}{D_{s}(\omega)}\right)^{2}
\frac{8i}{\pi d}\mu D_{0}\frac{1}{\pi N_{0}(0)}
\int\limits_{\omega}^{\infty}d\Omega\int \frac{d^{d}q}{(2\pi)^{d}}
\frac{q^{2}}{(-i(\Omega +\omega)+D(\Omega +\omega)q^{2})(-i\Omega+
D(\Omega)q^{2})^{2}}$$

{\bf Scheme III.}\ \ Now we can treat the Coulomb correction to relaxation
kernel $\delta M_{c}(\omega)$ in the right-hand side of self-consistency
equation (\ref{Ms}) also in self-consistent manner:

\begin{equation}
\delta M_{c}(\omega)=-\frac{M(\omega)}{D(\omega)}\delta D_{c}(\omega)
\label{dMf}
\end{equation}
In this case the equation for diffusion coefficient becomes:

\begin{equation}
\frac{D_{0}}{D(\omega)}=  1+ \frac{1}{ \pi N_{0}(0) }\int \
frac{d^{d}q}{(2\pi)^{d}} \frac{1}{-i\omega + D(\omega)q^2}-
\label{III}
\end{equation}
$$-\left(\frac{D_{0}}{D(\omega)}\right)^{2}
\frac{8i}{\pi d}\mu D_{0}\frac{1}{\pi N_{0}(0)}
\int\limits_{\omega}^{\infty}d\Omega\int \frac{d^{d}q}{(2\pi)^{d}}
\frac{q^{2}}{(-i(\Omega +\omega)+D(\Omega +\omega)q^{2})(-i\Omega+
D(\Omega)q^{2})^{2}}$$

{\bf Scheme IV.}\ \ As a zeroth approximation we can again take the usual
self-consistent theory of localization but self-consistency scheme can be
realized considering corrections to diffusion coefficient itself, not to the
relaxation kernel.

Self-consistent equation (\ref{ES}) can be rewritten as:

\begin{equation}
\frac{D(\omega)}{D_{0}}=\frac{1}{1+ \frac{1}{ \pi N_{0}(0) }\int
\frac{d^{d}q}{(2\pi)^{d}} \frac{1}{-i\omega + D(\omega)q^2}}
\label{Ds1}
\end{equation}
Let us now add to the right-hand side the correction to diffusion coefficient
Eq.(\ref{dD1}), so that finally we obtain the following equation:

\begin{equation}
\frac{D(\omega)}{D_{0}}=  \frac{1}{1+ \frac{1}{ \pi N_{0}(0) }\int
\frac{d^{d}q}{(2\pi)^{d}} \frac{1}{-i\omega + D(\omega)q^2}}+
\label{IV}
\end{equation}
$$+\frac{8i}{\pi d}\mu D_{0}\frac{1}{\pi N_{0}(0)}
\int\limits_{\omega}^{\infty}d\Omega\int \frac{d^{d}q}{(2\pi)^{d}}
\frac{q^{2}}{(-i(\Omega +\omega)+D(\Omega +\omega)q^{2})
(-i\Omega+D(\Omega)q^{2})^{2}}$$

The choice between alternative schemes I-IV is difficult to make on general
grounds. Thus, first of all, we have to analyze these schemes qualitatively.

\section{Qualitative analysis of different self-consistency schemes in
metallic phase.}

In the following we consider only three-dimensional systems. In self-consistent
theory of localization the diffusion coefficient is determined by
Eq.(\ref{Ds}).
{}From here it is easy to find the relaxation kernel:

\begin{equation}
\frac{M_{s}(\omega)}{M_{0}}=\frac{D_{0}}{D_{s}(\omega)}
\label{MDs}
\end{equation}
Consider now the correction to relaxation kernel which is given by the
following expression:

\begin{equation}
\frac{\delta M_{0}(\omega)}{M_{0}}=-\frac{\delta D_{c}(\omega)}{D_{0}}
\label{dMD}
\end{equation}
(As a zeroth approximation we take the usual Drude metal here.)

The Coulomb correction to diffusion coefficient of the usual self-consistent
theory of localization is determined by Eq.(\ref{dD}). Using this in Eq.
(\ref{dMD}), we obtain the appropriate correction to relaxation kernel
 $\delta M_{0s}(\omega)$.

Comparing $\delta M_{0s}(\omega)$ with $M_{s}(\omega)$, it is easy to see that
in metallic region and for the whole interval of frequencies we have the
following qualitative estimate:

\begin{equation}
\delta M_{0s}(\omega)\approx c\mu M_{s}(\omega)
\label{dMMs}
\end{equation}
where $c$ - is a numerical coefficient of the order of unity.

Detailed analysis shows that the similar estimate in metallic phase is valid
for relaxation kernel correction in all schemes discussed above. The reason
here is that all schemes of self-consistency in metallic phase the frequency
behavior of diffusion coefficient is similar to that of the usual
self-consistent theory of localization --- it is a constant for small
frequencies and  $D(\omega)\sim (-i\omega)^{1/3}$ for high-frequency region.
Thus, instead of Eq.(\ref{dMMs}) we can write:

\begin{equation}
\delta M_{0}(\omega)\approx c\mu M(\omega)
\label{dMM}
\end{equation}
Using the expression (\ref{dMM}) we can analyze the qualitative behavior of
different self-consistency schemes in metallic phase for $\omega=0$.

{\bf Scheme I.}\ \ We are considering the metallic phase, thus for
$\omega =0$ we have  $D(\omega =0)=D\neq 0$.
Multiplying Eq.(\ref{I}) by $M_{0}$, we obtain:
$$M=M_{0}+\frac{M_{0}}{ \pi N_{0}(0) }\int \frac{d^{d}q}{(2\pi)^{d}}
\frac{1}{Dq^{2}}+\delta M_{0}(0)$$
Using Eq.(\ref{dMM}), we get:
$$M=M_{0}+3\lambda x_{0}M+c\mu M$$
Accordingly $M=\frac{M_{0}}{1-3\lambda x_{0}-c\mu}$ and

\begin{equation}
\frac{D}{D_{0}}=1-3\lambda x_{0}-c\mu=\alpha -\alpha^{\ast}
\label{DI}
\end{equation}
where $\alpha^{\ast}=c\mu$.

Thus, even in the presence of weak Coulomb interaction continuous Anderson
transition persists and conductivity exponent remains the usual $\nu=1$
(i.e. diffusion coefficient linearly goes to zero with disorder parameter
$\alpha-\alpha^{\ast}$). However, we observe the shift of transition into the
region of more weak disorder $\alpha=\alpha^{\ast}$. Dependence of diffusion
coefficient on disorder is shown in Fig.3(a).

Note that if we analyze Eq.(\ref{I}) assuming the generalized diffusion
coefficient independent of frequency, we shall obtain discontinuous
metal-insulator transition with minimal metallic conductivity.

{\bf Scheme II.}\ \ Multiplying Eq.(\ref{II}) by $M_{0}$ and using
Eq.(\ref{dMM}) we obtain:
$$M=M_{0}+3\lambda x_{0}M+\frac{1}{\alpha^{2}}c\mu M$$
{}From here we have $M=\frac{M_{0}}{\alpha-c\mu/\alpha^{2}}$ or

\begin{equation}
\frac{D}{D_{0}}=\alpha-\frac{c\mu}{\alpha^{2}}
\label{DII}
\end{equation}

In this case again the continuous Anderson transition persists, conductivity
exponent is again $\nu=1$ and we only have the shift of the transition to
the region of weaker disorder $\alpha =(c\mu)^{1/3}$.

Dependence of diffusion coefficient on disorder is shown in Fig.3(b).

{\bf Scheme III.}\ \ Multiplying Eq.(\ref{III}) by $M_{0}$ and using
Eq.(\ref{dMM}), we obtain:
$$M=M_{0}+3\lambda x_{0}M+\left(\frac{M}{M_{0}}\right)^{2}c\mu M$$
or
$$\alpha\frac{M}{M_{0}}=1+c\mu \left(\frac{M}{M_{0}}\right)^{3}$$
In terms of diffusion coefficient:

\begin{equation}
\alpha=\frac{D}{D_{0}}+\frac{c\mu}{\left(\frac{D}{D_{0}}\right)^{2}}
\label{DIII}
\end{equation}
In this case we get the minimal metallic conductivity:
$$\frac{D_{min}}{D_{0}}=(2c\mu)^{1/3} \mbox{\ for\ } \alpha=
\alpha^{\ast}=\frac{3}{2}(2c\mu)^{1/3}$$
Dependence of diffusion coefficient on disorder is shown in Fig.3(c).

{\bf Scheme IV.}\ \ Diffusion coefficient correction is obtained from
Eq.(\ref{dMD}) using Eq.(\ref{dMM}):

\begin{equation}
\frac{\delta D_{c}(\omega)}{D_{0}}=-\frac{\delta M_{0}(\omega)}{M_{0}}
=-\frac{c\mu M(\omega)}{M_{0}}=-c\mu \frac{D_{0}}{D(\omega)}
\label{dDD}
\end{equation}
Substituting (\ref{dDD}) into Eq.(\ref{IV}) we get:
$$\frac{D}{D_{0}}=\frac{1}{1+(1-\alpha)\frac{D_{0}}{D}}-c\mu\frac{D_{0}}{D}$$
or
\begin{equation}
\alpha=\frac{c\mu}{\left(\frac{D}{D_{0}}\right)^{2}+c\mu}+\frac{D}{D_{0}}
\label{DIV}
\end{equation}
Again we obtain the minimal metallic conductivity:
$$\frac{D_{min}}{D_{0}}\approx (2c\mu)^{1/3} \mbox{\ for\ } \alpha
=\alpha^{\ast}\approx \frac{3}{2}(2c\mu)^{1/3}$$
Dependence of diffusion coefficient on disorder is shown in Fig.3(d).

\vskip 0.5cm

Qualitatively the results of schemes III and IV coincide and lead to the
concept of minimal metallic conductivity which does not correspond to the
majority of experimental data we have in this field [1, 2]. Schemes I and II
lead to the same continuous behavior of diffusion coefficient, differences
appear only in corresponding estimates of critical disorder at the transition.
It is quite natural that our approximations lead to the increased tendency
toward metal-insulator transition due to interelectron interactions. The
transition takes place at smaller disorders than in the absence of interaction.

In the following we shall concentrate on self-consistency scheme I where
Coulomb and localization corrections are being treated on the equal footing.

\section{Frequency dependence of diffusion coefficient.}

In self-consistency scheme I diffusion coefficient is defined by the integral
equation (\ref{I}). Let us transform it to dimensionless Matsubara frequencies:
 $ \frac{-i\omega}{D_{0}k_{0}^{2}}\rightarrow \omega,\ \frac{-i\Omega}
{D_{0}k_{0}^{2}}\rightarrow \Omega $,
and also introduce the dimensionless diffusion coefficient $d(\omega)=
\frac{D(\omega)}{D_{0}}$. In these notations integral equation (\ref{I})
takes the following form:

$$\frac{1}{d(\omega)}=  1+\frac{1}{d(\omega)}d\lambda x_{0}^{d-2}\int
\limits_{0}^{1}\frac{dyy^{d-1}}{y^{2}+\frac{\omega}{d(\omega)}}+$$
\begin{equation}
+\frac{8}{\pi}\mu\lambda x_{0}^{d-2}\int\limits_{\omega}^{\infty}
\frac{d\Omega}{d(\omega +\Omega)d^{2}(\Omega)}
\int\limits_{0}^{1}\frac{dyy^{d+1}}{\left(y^{2}+\frac{\omega+\Omega}
{d(\omega+\Omega)}\right)\left(y^{2}+\frac{\Omega}{d(\Omega)}\right)^{2}}
\label{In}
\end{equation}
In the following we shall limit ourselves only to the case of spatial dimension
d=3. Diffusion coefficient of the usual self-consistent theory of localization
(\ref{Ds}) in the same notations reduces to:

\begin{equation}
d(\omega)= \left\{
\begin{array}{lll}
\alpha=1-3\lambda x_{0} &
\quad \omega \ll \omega_{c}, \quad \alpha >0  & \quad \mbox{Metal}   \\
\left(\frac{\pi}{2}3\lambda x_{0}\right)^{\frac{2}{3}}\omega^{\frac{1}{3}} &
\quad \omega \gg \omega_{c} & \quad \mbox{Metal and Insulator} \\
\frac{\left(\frac{\pi}{2}3\lambda x_{0}\right)^{2}}{\alpha^{2}}\omega
=(\xi k_{0})^{2}\omega &
\quad \omega \ll \omega_{c}, \quad \alpha <0 & \quad \mbox{Insulator}
\end{array}\right.
\label{ds}
\end{equation}
where $\omega _{c}=\frac{|\alpha |^{3}}{\left(\frac{\pi}{2}3\lambda x_{0}
\right)^{2}}$ and $\xi$ - is the localization length.
Let us introduce $K(\omega)=\frac{\omega}{d(\omega)}$ and analyze Eq.(\ref{In})
assuming that $K(\omega)$, $K(\Omega)$ and $K(\omega +\Omega)\ll 1$.
Expanding the right-hand side of Eq.(\ref{In}) over these small parameters
we obtain:

$$\frac{\alpha}{d(\omega)}=1-\frac{\pi}{2}\frac{3\lambda x_{0}}{d(\omega)}
K^{1/2}(\omega)+$$
\begin{equation}
+2\mu\lambda x_{0}\int\limits_{\omega}^{\infty}\frac{d\Omega}{d(\omega
+\Omega)d^{2}(\Omega)}
\frac{K^{1/2}(\Omega)+2K^{1/2}(\omega+\Omega)}{\left(K^{1/2}(\Omega)
+K^{1/2}(\omega +\Omega)\right)^{2}}
\label{Inl}
\end{equation}
Consider the metallic phase and look for diffusion coefficient $d(\omega)$
solution in the following form:

\begin{equation}
d(\omega)= \left\{
\begin{array}{ll}
d & \quad \omega \ll \omega_{c} \\
d{\left(\frac{\omega}{\omega_{c}}\right)}^{\frac{1}{3}} & \quad \omega \gg
\omega_{c}
\end{array}\right.
\label{dm}
\end{equation}
Substituting (\ref{dm}) into Eq.(\ref{Inl}) we find $d$ and $\omega_{c}$
and for the diffusion coefficient we obtain:

\begin{equation}
d(\omega)= \left\{
\begin{array}{ll}
\alpha-\alpha^{\ast} & \quad \omega \ll \omega_{c} \\
\left(\frac{\pi}{2}3\lambda x_{0}\right)^{\frac{2}{3}}
\omega^{\frac{1}{3}} & \quad \omega \gg \omega_{c}
\end{array}\right.
\label{dml}
\end{equation}
where $\omega _{c}=\frac{|\alpha -\alpha^{\ast}|^{3}}{\left(\frac{\pi}{2}3
\lambda x_{0}\right)^{2}}$,
$\alpha ^{\ast}=c\mu $, $c\approx 0,89$

Thus for the metallic phase we confirmed the previous qualitative conclusion
---
Anderson transition persists and the conductivity exponent remains $\nu=1$.
The transition itself has shifted to the region of weaker disorder
$\alpha=\alpha^{\ast}=c\mu $.
The frequency behavior of diffusion coefficient in metallic phase is
qualitatively similar to that in the usual self-consistent theory of
localization (\ref{ds}). In the region of high frequencies $\omega \gg
\omega_{c}$ the behavior of diffusion coefficient remains unchanged after the
introduction of interelectron interactions.

Consider now the insulating phase. In the region of high frequencies
$\omega \gg \omega_{c}$ the diffusion coefficient obviously possess the
frequency dependence like $d(\omega) \sim \omega ^{1/3}$. Assume that for small
frequencies it is also some power of the frequency:

\begin{equation}
d(\omega)= \left\{
\begin{array}{ll}
d{\left(\frac{\omega}{\omega_{c}}\right)}^{\delta} & \quad \omega
\ll \omega_{c} \\
d{\left(\frac{\omega}{\omega_{c}}\right)}^{\frac{1}{3}} & \quad \omega
\gg \omega_{c}
\end{array}\right.
\label{dd}
\end{equation}
where $\delta$ is some exponent to be determined.

Substituting (\ref{dd}) into (\ref{Inl}) and considering the case of
$\alpha <0$ (insulating phase of the usual self-consistent theory of
localization) and $|\alpha |\gg \alpha^{\ast}$,\ we get:

\begin{equation}
d(\omega)= \left\{
\begin{array}{ll}
\frac{\left(\frac{\pi}{2}3\lambda x_{0}\right)^{2}}{\alpha^{2}}\omega
= (\xi k_{0})^{2}\omega &
\quad \omega ^{\ast}\ll \omega \ll \omega_{c} \\
\left(\frac{\pi}{2}3\lambda x_{0}\right)^{\frac{2}{3}}\omega^{\frac{1}{3}} &
\quad \omega \gg \omega_{c}
\end{array}\right.
\label{ddl}
\end{equation}
where  $\omega _{c}=\frac{|\alpha |^{3}}{\left(\frac{\pi}{2}3\lambda x_{0}
\right)^{2}}$,\ while $\omega ^{\ast}\approx 0,1\mu \frac{\alpha^{2}}
{\left(\frac{\pi}{2}3\lambda x_{0}\right)^{2}}=0,1\frac{\mu}{(\xi k_{0})^{2}}$
---is some new characteristic frequency defined by the interactions. Note that
$\omega ^{\ast}\rightarrow 0$ as we approach the transition point when
$\xi\rightarrow\infty$.

Thus, sufficiently deep inside the insulating phase when
$\alpha <0$ É $|\alpha |\gg \alpha^{\ast}$ and for the frequencies
$\omega \gg \omega ^{\ast}$, the diffusion coefficient remains the same as in
the self-consistent theory of localization,\ i.e. at small frequencies it is
linear over frequency,\ while for the higher frequencies it is
$\sim \omega^{1/3}$.

The analysis of Eq.(\ref{Inl}) shows that for the frequencies
$\omega \ll \omega ^{\ast}$ it is impossible to find the power-like dependence
for $d(\omega)$, i.e.\ the diffusion coefficient in the insulating phase is
apparently can not be represented in the form of
$d(\omega)=d\frac{\omega ^{\ast}}{\omega_{c}}{\left(\frac{\omega}
{\omega ^{\ast}}\right)}^{\delta}$,
where $\delta$ - is some unknown exponent. Because of this we were unable to
find any analytical treatment of Eq.(\ref{Inl}) in the region of
$\omega \ll \omega ^{\ast}$ within the insulating phase.

Consider now the system behavior not very deep inside the insulating phase
when  $\alpha -\alpha ^{\ast}<0$ while $\alpha >0$, that is when the system
without interaction would be within the metallic phase.

Let us assume that the frequency behavior of the diffusion coefficient for
$\omega\ll \omega _{c}$ possess the power-like form,\ i.e. the diffusion
coefficient is defined by the expression (\ref{dd}).

Substituting (\ref{dd}) into (\ref{Inl}) we get $\delta =\frac{1}{3}$.
As a result for the diffusion coefficient we get:

\begin{equation}
d(\omega)= \left\{
\begin{array}{ll}
\left(4,2\frac{\mu\lambda x_{0}}{\alpha}\right)^{\frac{2}{3}}
\omega^{\frac{1}{3}} &
\quad \omega \ll \omega_{c} \\
\left(\frac{\pi}{2}3\lambda x_{0}\right)^{\frac{2}{3}}\omega^{\frac{1}{3}} &
\quad \omega \gg \omega_{c}
\end{array}\right.
\label{dd0}
\end{equation}
where $\omega _{c}=\frac{|\alpha -\alpha ^{\ast}|^{3}}{\left(\frac{\pi}{2}3
\lambda x_{0}\right)^{2}}$.

Naturally, the exact solution for the diffusion coefficient should show
the continuous change of frequency behavior around  $\omega \sim \omega _{c}$.

Thus, within the insulating phase close to transition point,\ where the
system without interactions should have been metallic,\ the diffusion
coefficient behaves as $\sim \omega ^{1/3}$,\ everywhere,\ though for the low
frequency region the coefficient of $\omega ^{1/3}$ differs from that of
the usual self-consistent theory of localization and explicitly depends upon
the interaction constant.

Note that if we use the dynamically screened Coulomb interaction (\ref{V})
when $\mu\sim 1$,\ the region of applicability of (\ref{dd0}) widens,\ because
of $\alpha ^{\ast}\sim \mu\sim 1$.

We have also performed the numerical analysis of the integral equation
(\ref{In}) for the wide region of frequencies, both for metallic (Fig.4) and
insulating phases (fig.5). Numerical data are in good correspondence with our
analytical estimates.

In the region of high frequencies, both for metallic and insulating phases,
the frequency behavior of diffusion coefficient is very close to that defined
by the usual self-consistent theory of localization.

In the region of small frequencies within the metallic phase diffusion
coefficient $d(\omega)$ diminishes as interaction grows. Dependence of static
generalized diffusion coefficient on disorder for $\mu =0,24$ is shown at the
insert of Fig.4, and is practically linear. Metal-insulator transition in this
case is observed at $\alpha =\alpha^{\ast}= Ó\mu$, where $c\approx 0,5$,
which is also in good correspondence with our qualitative analysis.

Within the insulating phase for the region of small frequencies
($\omega \ll \omega^{\ast}$) we observe significant deviations from predictions
of the usual self-consistent theory of localization. Diffusion coefficient is
apparently nonanalytic in frequency here and we clearly see the tendency to
formation of some kind of effective gap for the frequencies $\omega \ll
\omega^{\ast}$.

Let us stress that our numerical analysis was performed in Matsubara frequency
region, which was used in writing down the Eq.(\ref{In}). Analytical
continuation of our numerical data to the real frequencies was not attempted.

\section
{Density of states close to the metal-insulator transition.}

Consider the effects of interelectron interactions upon the single-particle
("tunneling") density of states which is defined by the well-known relation:

\begin{equation}
N( \varepsilon )=-\frac{1}{\pi} \int \frac{ d^3 {\bf p} }{ (2\pi)^3 }
Im G^{R}({\bf p},\varepsilon)
\label{Ne}
\end{equation}
where $\varepsilon=E-E_{F}$ - is electronic energy with respect to the Fermi
level and $G^{R}({\bf p},\varepsilon)$ - is the retarded Green's function
defined by:

\begin{equation}
G^{R}({\bf p},\varepsilon) = \frac{1}{ \varepsilon -\xi_{{\bf p}}
+i\gamma-\Sigma^{R}_{ee}( \varepsilon, {\bf p}) }
\label{GG}
\end{equation}
Consider the so called "Fock" contribution to the self-energy part
$\Sigma^{R}_{ee}( \varepsilon, {\bf p})$ shown in Fig.6:

$$
\Sigma^{R}_{ee}( \varepsilon, {\bf p}) =  i \int \frac{ d^3 {\bf q} }
{ (2\pi)^3 } \int_{\varepsilon}^{1/\tau} \frac{ d\omega}{2\pi}
G_{0}^{A}({\bf p-q},\varepsilon-\omega) v({\bf q})
\gamma^{2}({\bf q},\omega) \approx
$$

\begin{equation}
\approx i \gamma^2 \mu  G_{0}^{A}({\bf p},\varepsilon) \left( f(\varepsilon,
\omega_{c}) + ig(\varepsilon, \omega_{c}) \right),
\label{SIGF}
\end{equation}
where  $ f_{\varepsilon,\omega_{c}}$  and $g_{\varepsilon,\omega_{c}}$ are
defined by the relations:

\begin{equation}
f_{\varepsilon,\omega_{c}}=4 N^{-1}_{0}(0) Re \int \frac{ d^3 {\bf q} }
{ (2\pi)^3 } \int_{\varepsilon}^{1/\tau} \frac{ d\omega}{2\pi}\frac{1}
{\left( -i\omega + D_{E}(\omega)q^2 \right)^2}
\label{F}
\end{equation}

\begin{equation}
g_{\varepsilon,\omega_{c}}=4 N^{-1}_{0}(0) Im \int \frac{ d^3 {\bf q} }
{ (2\pi)^3 } \int_{\varepsilon}^{1/\tau} \frac{ d\omega}{2\pi}\frac{1}
{\left( -i\omega + D_{E}(\omega)q^2 \right)^2}
\end{equation}
and $G_{0}^{A}({\bf p},\varepsilon)$ - is the advanced Green's function
without interaction contributions. Here and in the following we denote as
$N_{0}(0)$ the density of states at the Fermi level in the absence of
interactions.

Substituting the expression for $\Sigma^{R}_{ee}( \varepsilon, {\bf p})$
from (\ref{SIGF}) into (\ref{GG}) we obtain as $ \varepsilon \rightarrow 0 $:

$$
\frac{N( \varepsilon )}{N_{0}(0)} \approx
-\frac{1}{\pi} \int \limits_{-\infty}^{\infty} d\xi_{{\bf p}}Im G^{R}({\bf p},
\varepsilon)=
$$
$$
=\frac{\gamma}{\pi} \int \limits_{-\infty}^{\infty} d\xi_{{\bf p}}
\frac{\xi^2_{{\bf p}}+\gamma^2+\mu\gamma^2g_{\varepsilon, \omega_{c}}}{\left(
\xi^2_{{\bf p}}+\gamma^2+\mu\gamma^2g_{\varepsilon, \omega_{c}}\right)^2
+\left( \mu\gamma^2f_{\varepsilon, \omega_{c}}\right)^2}=
$$
\begin{equation}
=\frac{1}{\sqrt{2(1+\mu g_{\varepsilon, \omega_{c}} +((1+\mu g_{\varepsilon,
\omega_{c}})^2 + (\mu f_{\varepsilon, \omega_{c}})^2  )^{1/2} )}}
\left(1 + \frac{1+\mu g_{\varepsilon, \omega_{c}}}{((1+\mu g_{\varepsilon,
\omega_{c}})^2 + (\mu f_{\varepsilon, \omega_{c}})^2  )^{1/2}}  \right)
\label{PL}
\end{equation}
Now let us calculate $f_{\varepsilon, \omega_{c}}$ É $g_{\varepsilon,
\omega_{c}}$.
As we noted above, the frequency behavior of diffusion coefficient in metallic
phase is qualitatively similar to that in the usual self-consistent theory of
localization (with the shifted mobility edge). In insulating phase, in a narrow
frequency region $\omega \ll \omega^{*}$ the frequency dependence of
$D(\omega)$ is actually unknown to us. However, taking into account the fact
that as we approach the transition $ \omega^{*} \rightarrow 0$,\ it is
reasonable to use the expression (5) for the generalized diffusion coefficient
$D(\omega)$,\ as given by the self-consistent theory of localization. Here we
assume that the role of interactions reduces only to a simple shift of
transition point.

In metallic region, for $ \varepsilon \rightarrow 0 $  we obtain:

\begin{equation}
f_{\varepsilon,\omega_{c}}=\frac{3^{3/2}}{\sqrt{2}} \left(\frac{\gamma}
{E_{F}} \right)^2 \left( 1-\frac{{\varepsilon}^{1/2}}{{\omega_{c}}^{1/2}}
\right) +  \frac{3^3}{\pi} \left( \frac{\gamma}{E_{F}} \right)^3
\left( \frac{\omega_{c}^{1/3}}{(1/\tau)^{1/3}} - 1 \right)
\label{f1}
\end{equation}

\begin{equation}
g_{\varepsilon,\omega_{c}}=\frac{3^{3/2}}{\sqrt{2}} \left(\frac{\gamma}
{E_{F}} \right)^2 \left( 1-\frac{{\varepsilon}^{1/2}}{{\omega_{c}}^{1/2}}
\right) +  \frac{3^{3/2}}{2} \left( \frac{\gamma}{E_{F}} \right)^2
ln \frac{1/\tau}{\omega_{c}}
\label{g1}
\end{equation}
In the region of $\omega_{c} < \varepsilon < 1/\tau$ the functions
$f_{\varepsilon,\omega_{c}}$ and $g_{\varepsilon,\omega_{c}}$ reduce to:

\begin{equation}
f_{\varepsilon,\omega_{c}}= \frac{3^3}{\pi} \left( \frac{\gamma}{E_{F}}
\right)^3
\left( \frac{\varepsilon^{1/3}}{(1/\tau)^{1/3}} - 1 \right)
\label{f2}
\end{equation}

\begin{equation}
g_{\varepsilon,\omega_{c}}= \frac{3^{3/2}}{2} \left( \frac{\gamma}{E_{F}}
\right)^2
ln \frac{1/\tau}{\varepsilon}
\label{g2}
\end{equation}
Using the expressions (\ref{f1}-\ref{g2}) we can analyze Eq.(\ref{PL}) for the
density of states $N(\varepsilon)$. In case of relatively weak disorder
$f_{\varepsilon,\omega_{c}}$ and $g_{\varepsilon,\omega_{c}} \ll 1$ and for
$\mu \ll 1$ we can restrict ourselves by the linear term over $\mu$ and obtain
the correction found in Refs.[10-11]:

\begin{equation}
\frac{N(\varepsilon)}{N_{0}(0)} \approx
\left( 1-\frac{\mu}{2} g_{\varepsilon,\omega_{c}}  \right)
\label{Napprox}
\end{equation}
As the system moves towards the mobility edge
($\gamma \sim E_{F}$, $\omega_{c} \rightarrow 0 $) the function
$g_{\varepsilon,\omega_{c}} $  diverges logarithmically
($g_{\varepsilon,\omega_{c}} \sim ln \frac{1/\tau}{\omega_{c}}$)
and the use of expression (\ref{Napprox}) is insufficient.
We have to calculate the density of states using the complete expression
(\ref{PL}). Note that in Refs.[10-11] the frequency dependence of diffusion
coefficient was completely neglected. However, we shall see that it will
become very important as we approach the transition.
For $\omega_{c} \rightarrow 0 $ in Eq.(\ref{PL}) we can neglect
$f_{\varepsilon,\omega_{c}}$ in comparison with the diverging function
$g_{\varepsilon,\omega_{c}}$. Thus we have:

\begin{equation}
\frac{N(\varepsilon)}{N_{0}(0)} \approx \left( 1+\mu g_{\varepsilon,
\omega_{c}}  \right)^{-1/2}
\approx 1 - \mu \frac{3^{3/2}}{2^{3/2}} \left( \frac{\gamma}{E_{F}} \right)^2
\left( 1 - \frac{\varepsilon^{1/2}}{\omega_{c}^{1/2}}
+ \frac{1}{\sqrt{2}} ln \frac{1/\tau}{\omega_{c}} \right)
\end{equation}
In case of the weak disorder the well-known square root singularity appears
at the Fermi level [3]. As system moves towards the mobility edge
($\omega_{c} \rightarrow 0 $) the divergence of $ln \frac{1/\tau}{\omega_{c}}$
causes this minimum to become more deep and at the mobility
edge itself ($\omega_{c} = 0 $) the density of states at the Fermi level
is equal to zero. Note that the square root dependence of the density of states
persists only in the region of $0 < \varepsilon < \omega_{c} $, and the width
of this region tends to zero as we move to the mobility edge. In the region of
$\omega_{c} < \varepsilon < 1/\tau $ we have:

\begin{equation}
\frac{N(\varepsilon)}{N_{0}(0)} \approx \left( 1+\mu \frac{3^{3/2}}{2}
\left(\frac{\gamma}{E_{F}}\right)^2 ln \frac{1/\tau}{\varepsilon}
\right)^{-1/2}
\label{mob}
\end{equation}
while at the mobility edge itself ($\gamma \sim E_{F}$):

\begin{equation}
\frac{N(\varepsilon)}{N_{0}(0)} \approx \left( 1+\mu \frac{3^{3/2}}{2}
ln \frac{1/\tau}{\varepsilon} \right)^{-1/2}
\end{equation}
In Fig.7 (curves 1-3) we show the numerical data for the density of states
obtained from direct calculations of
$f_{\varepsilon,\omega_{c}}$ É $g_{\varepsilon,\omega_{c}}$
and Eq.(\ref{PL}) which demonstrate the formation and growth of effective
pseudogap. Curves are given for the different values of parameter
$\lambda/\lambda_{c}$, where  $\lambda_{c}$ - is the critical disorder
parameter determining the metal-insulator transition. Dashed curve shows
the behavior of the density of states at the mobility edge itself
($\lambda/\lambda_{c}=1$).

In insulating phase the functions
$g_{\varepsilon,\omega_{c}}$ and $f_{\varepsilon,\omega_{c}}$
have the following form as $\varepsilon \rightarrow 0 $:

\begin{equation}
f_{\varepsilon,\omega_{c}}=\frac{3^{3/2}}{2} \left(\frac{\gamma}{E_{F}}
\right)^2 \left( 1-\frac{\omega_{c}}{\varepsilon} \right)
+ \frac{3^3}{\pi} \left(\frac{\gamma}{E_{F}} \right)^3
\left( \frac{\omega_{c}^{1/3}}{(1/\tau)^{1/3}} - 1 \right)
\label{f3}
\end{equation}

\begin{equation}
g_{\varepsilon,\omega_{c}}= \frac{3^{3/2}}{2} \left( \frac{\gamma}{E_{F}}
\right)^2
ln \frac{1/\tau}{\omega_{c}}
\end{equation}
Neglecting in Eq.(\ref{PL}) $g_{\varepsilon,\omega_{c}}$ in comparison with
$f_{\varepsilon,\omega_{c}}$ which diverges as $\varepsilon \rightarrow 0 $
we get:

\begin{equation}
\frac{N(\varepsilon)}{N_{0}(0)}
\approx \frac{1}{\mu^{1/2}} \frac{2^2}{3^{3/2}} \frac{\varepsilon^{1/2}}
{\omega_{c}^{1/2}}
\label{PL1}
\end{equation}
Thus, in the insulating region the square root dependence for
$\varepsilon \rightarrow 0 $ persists and its region widens as disorder
grows. For $\omega_{c} < \varepsilon < 1/\tau $ we obtain the previous
dependence given in Eq.(\ref{mob}) (Fig.7, curves 4-5).

Above we have considered the conduction band of infinite width. Important
changes for the insulating phase appear as we address the case of conduction
band of finite width $2÷$. In the model of linear electronic spectrum we
obtain:

\begin{equation}
\frac{N(\varepsilon)}{N_{0}(0)}=\frac{\gamma}{\pi} \int \limits_{-B}^{B}
d\xi_{{\bf p}}
\frac{\xi^2_{{\bf p}}+\gamma^2}{\left( \xi^2_{{\bf p}}+\gamma^2 \right)^2
+\left( \mu\gamma^2 f_{\varepsilon, \omega_{c}}\right)^2}
\label{DS}
\end{equation}
Analysis of expressions (\ref{f3}) and (\ref{DS}) shows that the density
of states in the region of $\varepsilon \ll \mu \omega_{c} \left(
\frac{\gamma}{B} \right)^2$ demonstrates quadratic $\varepsilon$ - behavior:

\begin{equation}
\frac{N(\varepsilon)}{N_{0}(0)} \sim \left( \frac{B}{\gamma} \right)^3
\left( \frac{E_{F}}{\gamma}\right)^4 \frac{1}{\mu^2}
\left( \frac{\varepsilon}{\omega_{c}}  \right)^2,
\end{equation}
which reminds the well known "Coulomb gap" of Efros and Shklovskii [13-14]
valid deep within insulating phase. Note that this may a pure coincidence
because the analysis of Efros and Shklovskii is essentially based upon the
long-range nature of Coulomb interaction. The numerical results for
$N(\varepsilon)$ are shown in Fig.8 (curves 5-6).
In metallic region in the case of conduction band of finite width numerical
data demonstrate that the density of states is practically unchanged in
comparison with the case of the infinite band.

Tunneling densities of states dependences similar to those shown in Fig.8
were observed experimentally for a number of disordered systems close to the
metal-insulator transition [1,2], from amorphous alloys [15-17] to
disordered single-crystals of metallic oxides, including high-temperature
superconductors [18]. While we see the complete qualitative correspondence of
theory and experiments, though we must stress that the quantitative agreement
is obviously absent --- the effective width of the "pseudogap" close to the
transition is larger in the experiments and can be described by power-like,
not by logarithmic dependence. At the same time our expressions describe the
significant overall drop of the density of states in rather wide energy
region, which is observed in all experiments. The expressions obtained above
give, as far as we know, the first description of the evolution of the
tunneling density of states during the transition from metallic to insulating
state induced by disordering, for the whole interval of parameters
controlling this transition.

\section{Conclusions.}

In this paper we propose a self-consistency scheme which allows the description
of disorder induced metal-insulator transition taking into account the
first-order effects of interelectron interactions. This approach is based upon
the assumption that neglecting the interelectron interactions this transition
can be sufficiently well described by self-consistent theory of localization,
while interaction effects may be taken into account if we consider the
appropriate interaction and localization diagrams on the equal footing in the
equation for the generalized diffusion coefficient.
We demonstrate that this approach allows us to formulate unified equations
describing the transition from metallic to insulating state and obtain a
number of results which are in qualitative agreement with experiments on
disordered systems (tunneling density of states). Interaction effects
remain relatively small in this approach and for the wide region of parameters
the diffusion coefficient is well described by self-consistent theory of
localization. In particular, metallic conductivity (for $T=0$) linearly drops
to zero at the transition, which is observed in a number of real systems [1,2]
However, this approach is unable to describe the known class of disordered
systems where this conductivity drop is apparently described by the square root
dependence on disorder parameter [1,2].

{}From purely theoretical point of view, even if we restrict ourselves only to
first-order interaction corrections, we still need to perform a complete
analysis of "Hartree"-type interaction corrections neglected above, as well as
the importance of screening anomalies in insulating phase [5,11]. We have
also neglected all the effects due to electronic spin, while the growing
importance of these effects as we approach the metal-insulator transition
were noted long ago [1,2].

Still, the proposed scheme of analysis is, in our opinion, of considerable
interest because it allows us to take into account interaction effects in a
wide region of parameters controlling the metal-insulator transition, including
even the insulating phase, which was inaccessible for the previous theories.

This work was partially supported by Russian Foundation for Fundamental
Research under the grant $N^{o}$ 93-02-2066 as well as by
Soros International Science Foundation grant RGL000.

\newpage
\setlength{\textwidth}{16.5cm}

\newpage
\setlength{\textwidth}{15.0cm}
\begin{center}
{\bf Figure Captions.}
\end{center}
\vskip 0.5cm

Fig.1. Two-particle Green's function $\Phi ^{RA}_{{\bf p p'}}({\bf q}\omega
\varepsilon)$
and vertex parts $\gamma^{RA}({\bf pq}\omega)$ and
$\Gamma^{RA}_{{\bf pp'}}({\bf q}\omega)$.

\vskip 0.3cm

Fig.2. First-order interaction corrections to conductivity.

\vskip 0.3cm

Fig.3. Static generalized diffusion coefficient ($d=\frac{D(0)}{D_{0}}$)
dependence on disorder in different self-consistency schemes:
(a) --- scheme I; (b) --- II; (c) --- III; (d) --- IV.

\vskip 0.3cm

Fig.4. Dependence of dimensionless generalized diffusion coefficient on
dimensionless Matsubara frequency in metallic phase ($\alpha =0.5$),
obtained by numerical solution of Eq.(\ref{In}) for different values of
$\mu$: {\bf 1.} 0,24; {\bf 2.} 0,6; {\bf 3.} 0,95;
Dashed line --- the usual self-consistent theory of localization, $\mu =0$.

At the insert: Dependence of static diffusion coefficient
($d=\frac{D(0)}{D_{0}}$) on disorder for $\mu =0,24$.

\vskip 0.3cm

Fig.5. Dependence of dimensionless generalized diffusion coefficient on
dimensionless Matsubara frequency in dielectric phase ($\alpha =-0.5$),
obtained by numerical solution of Eq.(\ref{In}) for different values of
$\mu$: {\bf 1.} 0,12; {\bf 2.} 0,6; {\bf 3.} 1,2;
Dashed line --- the usual self-consistent theory of localization, $\mu =0$.

\vskip 0.3cm

Fig.6. "Fock" contribution to self-energy part.

\vskip 0.3cm

Fig.7. Evolution of the density of states during the metal-insulator
transition in the interacting system with $\mu=0.2$. Infinitely wide
conduction band. Data are shown for different values of disorder parameter
$\lambda/\lambda_{c}$: ${\bf1.}$0.5; ${\bf2.}$0.7; ${\bf3.}$1;
${\bf4.}$1.4; ${\bf5.}$1.8

\vskip 0.3cm

Fig.8. Evolution of the density of states during metal-insulator transition
in the interacting system with $\mu=0.3$. Conduction band of finite width
$2B=4E_{F}$. Data are shown for different values of disorder parameter
$\lambda/\lambda_{c}$: ${\bf1.}$0.5; ${\bf2.}$0.7; ${\bf3.}$0.9;
${\bf4.}$1; ${\bf5.}$2; ${\bf6.}$2.4

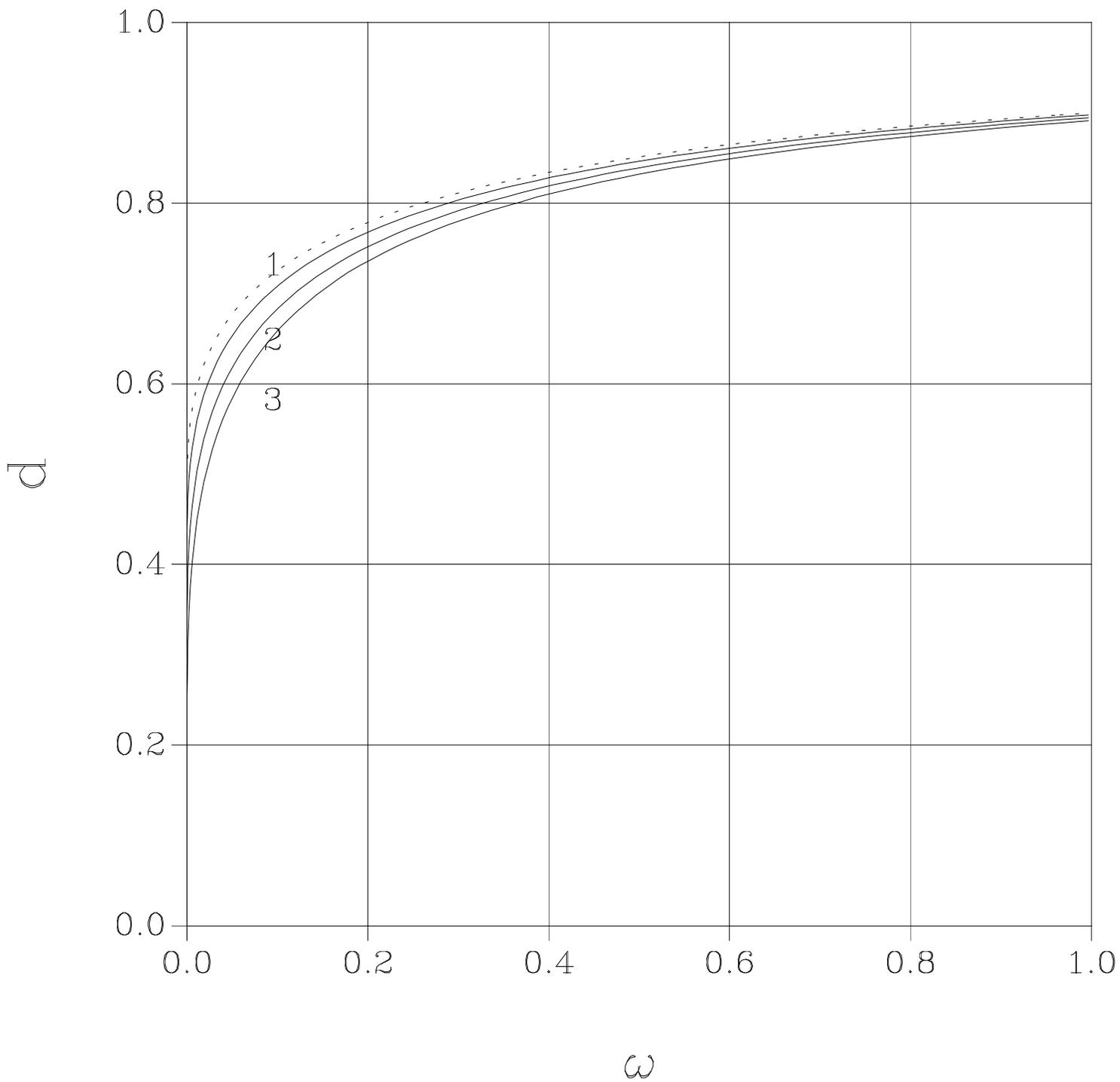
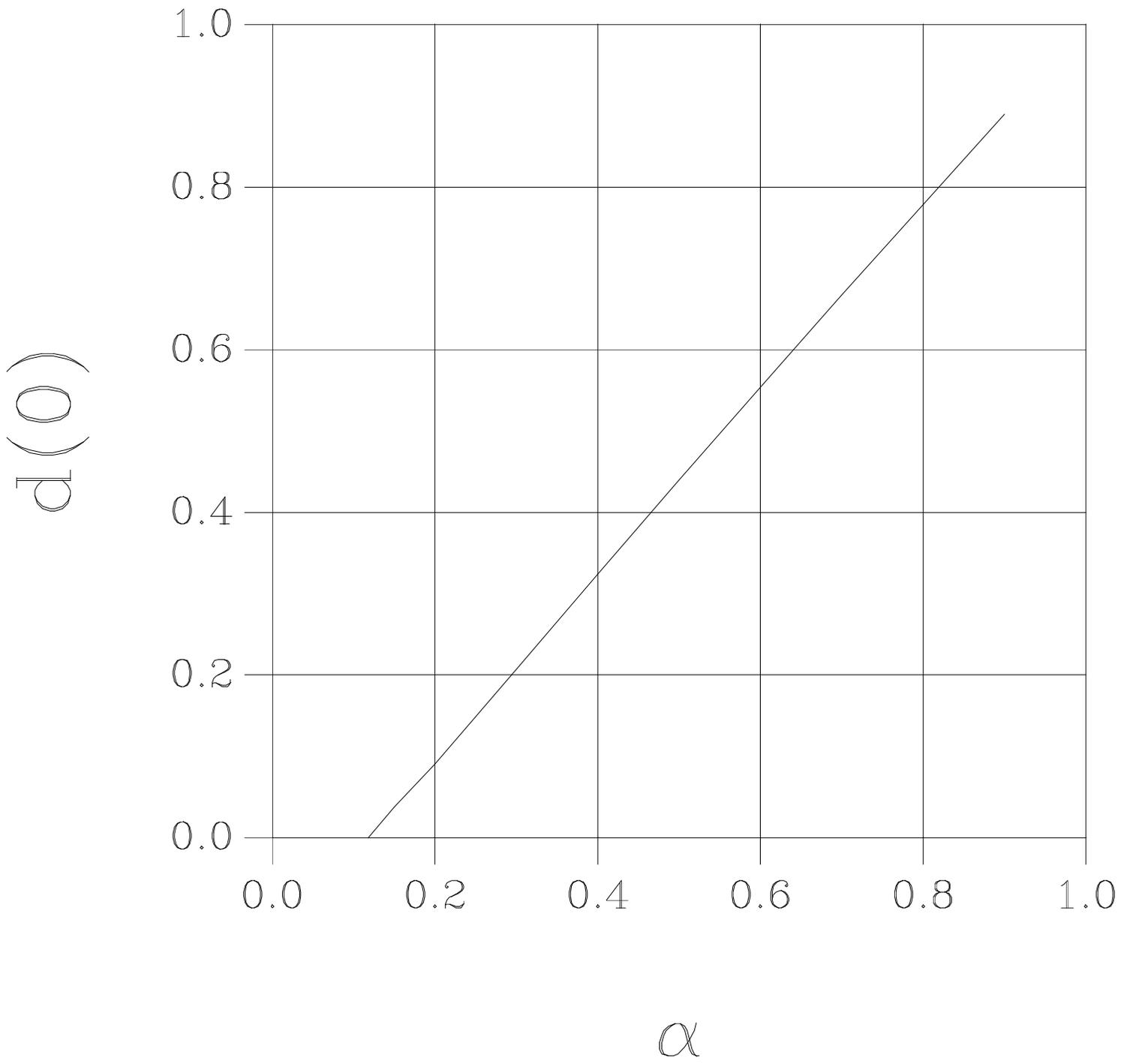
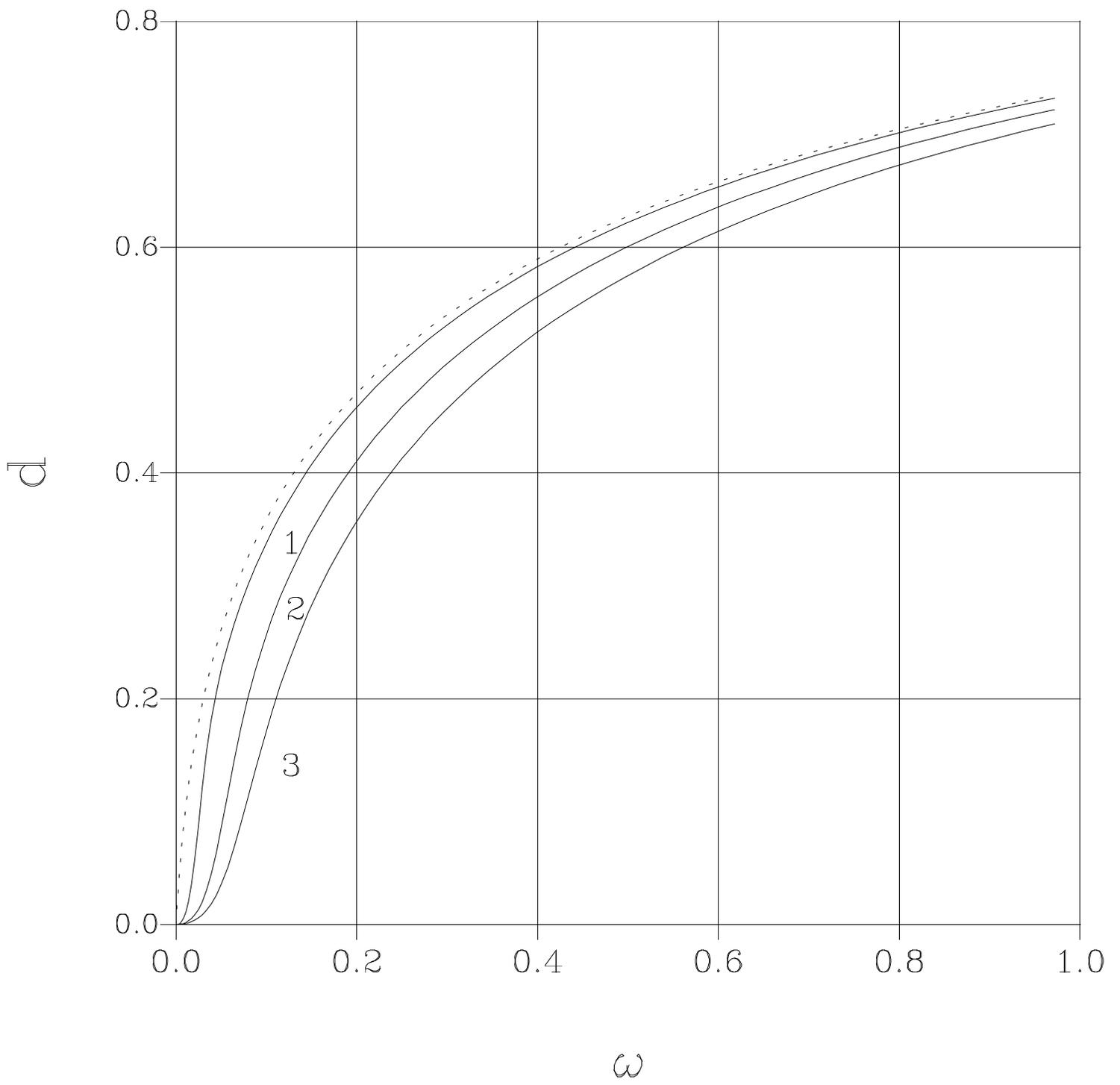
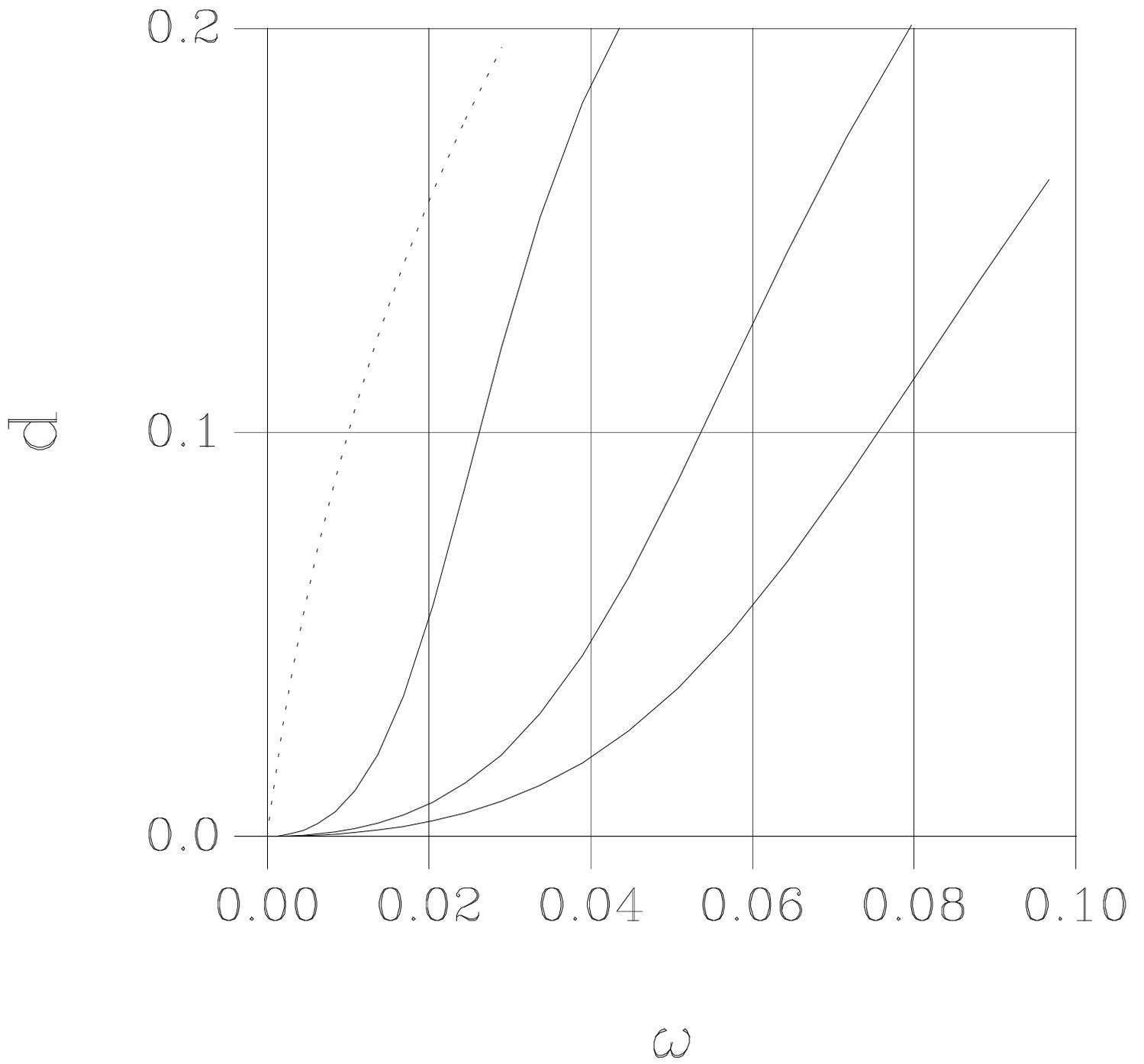
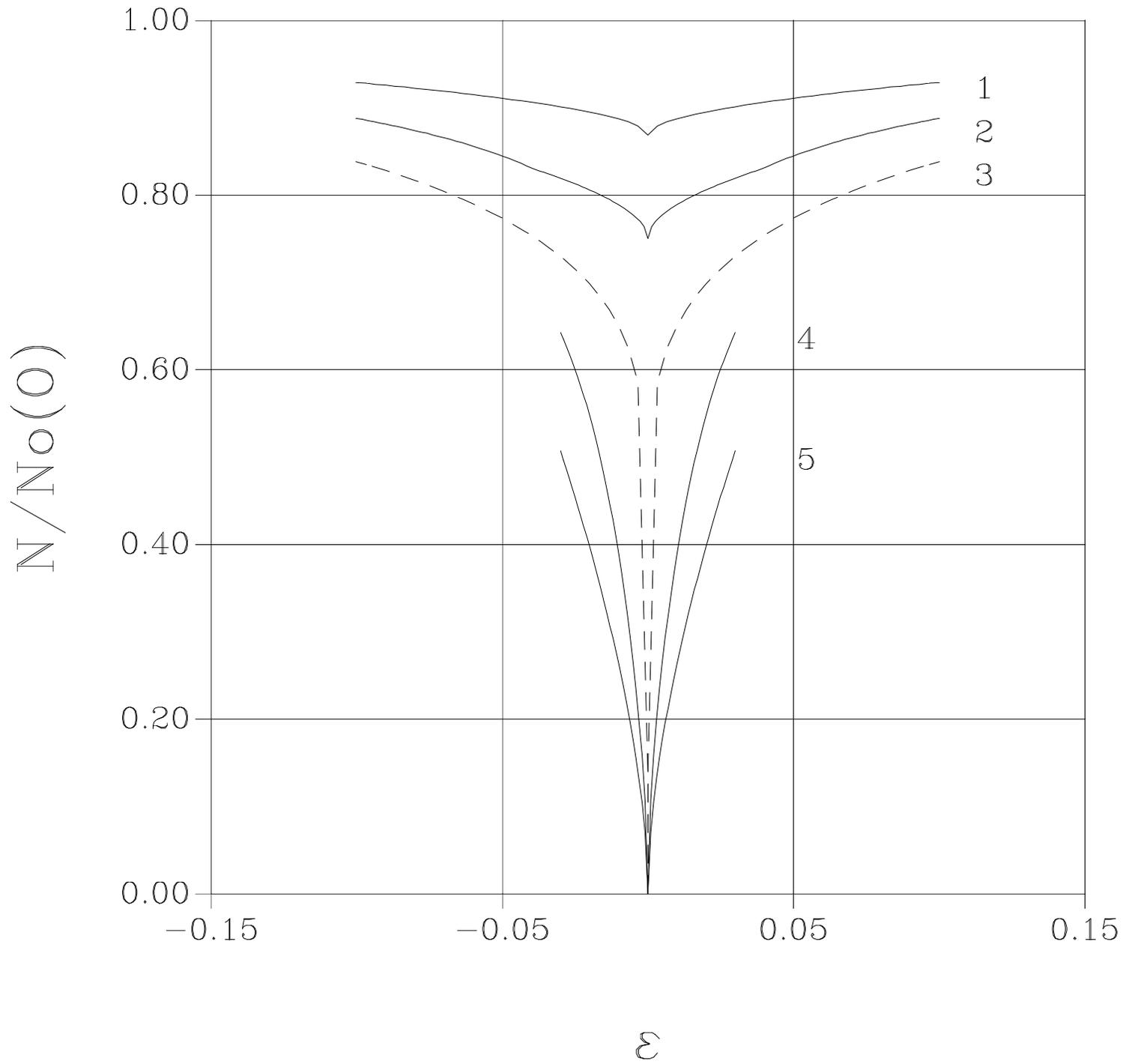
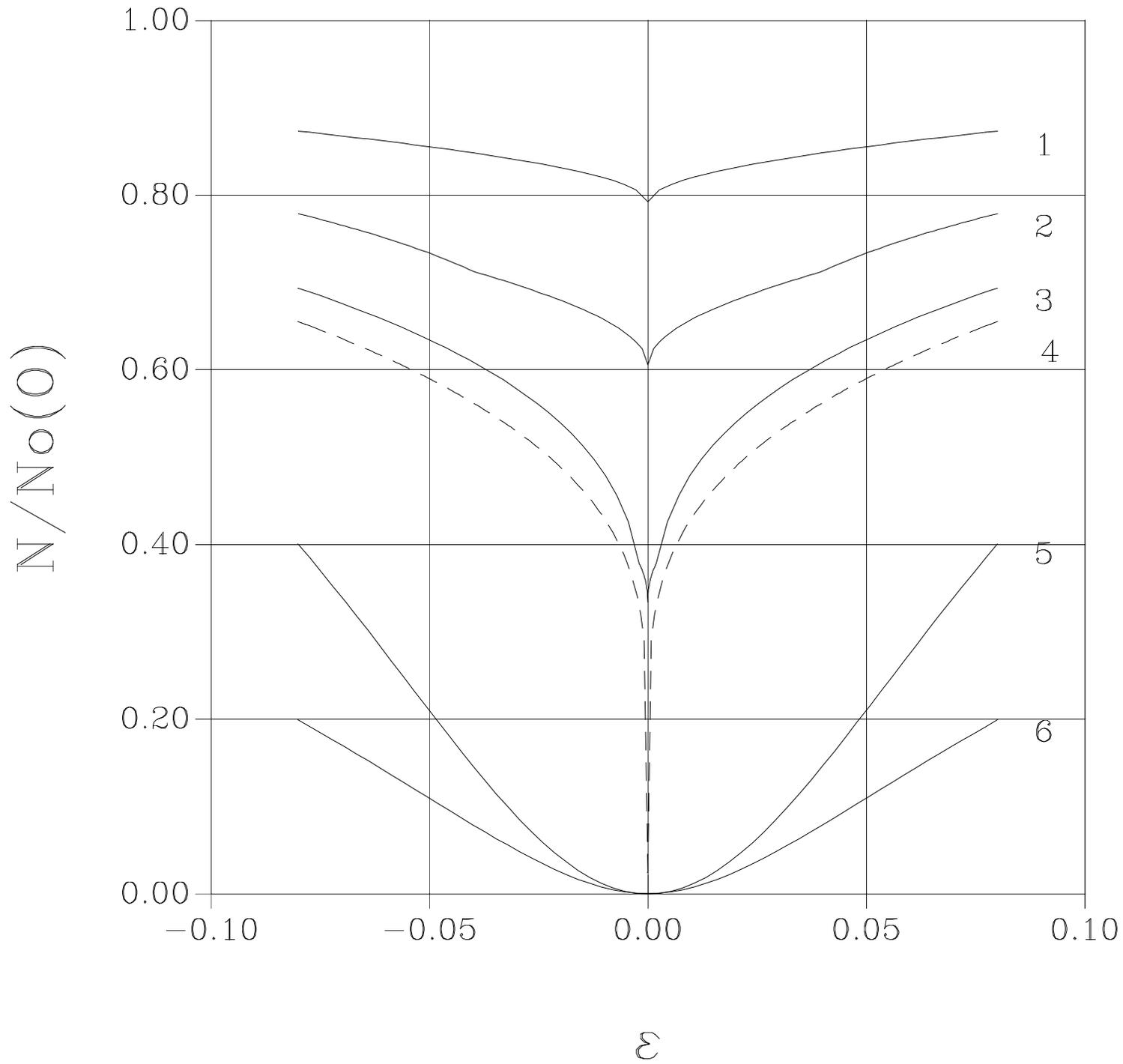
\end{document}